\documentclass[a4paper]{article}
\usepackage{amsmath,pstricks}
\newcommand{\scr}{\scriptscriptstyle}
\newcommand{\eqa}{\begin{eqnarray}}
\newcommand{\neqa}{\end{eqnarray}}
\newcommand{\equ}{\begin{equation}}
\newcommand{\nequ}{\end{equation}}

\begin{document}

\title{\bf{\LARGE Tensorial Structure \\ of the LQG graviton propagator}
}

\author{\large
Emanuele Alesci${}^{ab}$
 \\[3mm]
\em
\normalsize
{${}^a$Dipartimento di Fisica Universit\`a di Roma Tre, I-00146 Roma EU}
\\ \em\normalsize{${}^b$Centre de Physique Th\'eorique de Luminy%
\footnote{Unit\'e mixte de recherche (UMR 6207) du CNRS et des
Universit\'es de Provence (Aix-Marseille I), de la Mediterran\'ee
(Aix-Marseille II) et du Sud (Toulon-Var); laboratoire affili\'e \`a
la FRUMAM (FR 2291).}, Universit\'e de la M\'editerran\'ee, F-13288
Marseille EU}}

\date{\small\today}
\maketitle \vspace{-.6cm}
\maketitle

\begin{abstract}
We review the construction of the tensorial structure of the graviton propagator in the context of loop quantum gravity and spinfoam formalism. The main result of this analysis is that applying the same strategy used to compute the diagonal terms, the Barrett-Crane vertex is unable to yield the correct propagator in the long distance limit. The problem is in the intertwiner-independence of the Barrett-Crane vertex. We also review the asymptotic behavior of an alternative vertex that is able to give the correct propagator. 
\end{abstract}

\section{The LQG Graviton Propagator}
Loop quantum gravity (LQG) \cite{lqg} is one of the main candidates for a theory of quantum gravity. However LQG has difficulties with the low energy limit of the theory and the possibility to calculate scattering amplitudes.
A strategy for addressing those problems, based on the boundary formulation \cite{boundary} of n--points functions, applied to the calculation of the graviton propagator, has been introduced\cite{scattering1}  and developed\cite{Livine:2006it}.
Here we review the main achievements in the construction of the euclidean graviton propagator $G^{\mu\nu\rho\sigma}(x,y)$ in the context of LQG. If we choose a regular 4-simplex with two boundary tetrahedra $n$ and $m$ centered at the points $x$ and $y$ we can define ${\mathbf G} _{n,m}^{\scriptscriptstyle ij,kl}(L)=
G^{\mu\nu\rho\sigma}(x,y)(n^{\scriptscriptstyle(i)}_n)_\mu (n^{\scriptscriptstyle(j)}_n)_\nu (n^{\scriptscriptstyle(k)}_m)_\rho (n^{\scriptscriptstyle(l)}_m)_\sigma$, where the latin indexes label the five tetrahedra bounding the 4-simplex and $n_m^{\scriptscriptstyle(k)}$ is the normal one-form to the triangle bounding the tetrahedra $m$ and $k$,  in the hyperplane defined by $m$, and $L$ is the euclidean distance between $x$ and $y$. Knowing  ${\mathbf G} _{n,m}^{\scriptscriptstyle ij,kl}(L)$ is the same as knowing $G^{\mu\nu\rho\sigma}(x,y)$.   ${\mathbf G} _{n,m}^{\scriptscriptstyle ij,kl}(L)$  can be computed\cite{scattering1}
in a background  independent context as 
\begin{equation}
{\mathbf G} _{{\mathbf q}\, n,m}^{\scriptscriptstyle ij,kl} = \langle W | \big(E^{\scriptscriptstyle(i)}_n \cdot E^{\scriptscriptstyle(j)}_n-n_n^{\scriptscriptstyle(i)}\cdot n_n^{\scriptscriptstyle(j)}\big)
\big(E^{\scriptscriptstyle(k)}_m  \cdot E^{\scriptscriptstyle(l)}_m-n_m^{\scriptscriptstyle(k)}\cdot n^{\scriptscriptstyle(l)}_m\big) |\Psi_{\mathbf q} \rangle.
\label{partenza1}
\end{equation}
for an appropriate $\mathbf q$. We refer to \eqref{partenza1} as the LQG graviton propagator.  Here $ \langle W |$ is the boundary functional.   The operator $E_n^{\scriptscriptstyle(i)}$ is the triad operator at the point $n$, contracted with $n_n^{\scriptscriptstyle(i)}$.  $|\Psi_{\mathbf q} \rangle$
is the boundary state, picked on a given classical boundary (intrinsic and 
extrinsic)  geometry $\mathbf q$. 
The diagonal components ${\mathbf G} _{{\mathbf q}\, n,m}^{\scriptscriptstyle ii,kk}$
were computed in Ref. \cite{scattering1}.
Using a gaussian form of the vacuum state and the Barret Crane \cite{BC}(BC) dynamics the expression\cite{scattering1} of the diagonal components  \emph{ at large distance, agrees with the conventional graviton propagator}! 
The next natural step was the reconstruction of the whole tensorial structure of the LQG propagator. This analysis has been performed in the articles Ref.\cite{I} and Ref. \cite{II}.

The construction of the non diagonal terms requires to think over the whole used theory, because the graviton operators $E^{\scriptscriptstyle(i)}_n \cdot E^{\scriptscriptstyle(j)}_n$ call into play the dependence of the spinnetworks from the \emph {intertwiners} and in turns, the dependence of the boundary state and the vertex from these variables.
In particular the BC dynamics used to compute the diagonal terms has a trivial intertwiner dependence that appear insufficient to deal with the non diagonal terms.
In Ref \cite{I} the authors find that the BC 
vertex \emph{fails} to give the correct propagator in the large-distance limit. 
In Ref. \cite{II} is presented the asymptotic behavior of a vertex amplitude $W$ that yields the correct propagator.
%The paper is organized as follows in the next section we give the basic ingredients to compute the LQG graviton propagator, in Section 3 we summarize the problems of the BC models in the construction of the graviton propagator and in Section 4 we give an example of an alternative vertex that yield the correct propagator.
%\section{Propagator Ingredients}

Eq. \eqref{partenza1} to first order in the GFT\cite{lqg,gft} expansion, and in the limit in which the boundary surface is large receive the leading contribution for $W$ with support only on spin networks with a 4-simplex graph. If 
${\mathbf j}=(j_{nm})$ and ${\mathbf i}=(i_n)$
are, respectively, the ten spins and the five intertwiners that color this graph, then in this
approximation (\ref{partenza1}) reads
$
{\mathbf G} _{{\mathbf q}\, n,m}^{\scriptscriptstyle ij,kl} 
 = \sum{}_{{\mathbf j}, {\mathbf i}} \,W({\mathbf j}, {\mathbf i}) \big(E^{\scriptscriptstyle(i)}_n \cdot E^{\scriptscriptstyle(j)}_n-n_n^{\scriptscriptstyle(i)}\cdot n_n^{\scriptscriptstyle(j)}\big) 
 \big(E^{\scriptscriptstyle(k)}_m  \cdot E^{\scriptscriptstyle(l)}_m-n_m^{\scriptscriptstyle(k)}\cdot n^{\scriptscriptstyle(l)}_m\big)
 \Psi({\mathbf j}, {\mathbf i}).
$
The calculation of this expression requires the use of three ingredients: the double grasping operators, the boundary state and the vertex amplitude of a 4--simplex.

%\subsection{Operators}

The action of the double grasping operators was computed in Ref. \cite{I}. They act on a 4--valent node in four possible ways.  
The diagonal action (the only used in Ref. \cite{scattering1}) is
$	E_n^{\scr(i)} \cdot E_n^{\scr(i)}\left|{\mathbf j},{\mathbf i}\right\rangle=C^{\scr{ni}}\left|{\mathbf j},{\mathbf i}\right\rangle
$
where $C^{\scr{ni}}$ is the Casimir of the irrep. associated to the link $ni$. 
The non-diagonal action give a diagonal operator  
$
	E_n^{\scr(i)} \cdot E_n^{\scr(j)}\left|{\mathbf j},{\mathbf i}\right\rangle
=\sum_{\mathbf j,\mathbf i}
D^{ij}_{n}\ \left|{\mathbf j},{\mathbf i}\right\rangle,
$
and two possible non-diagonal operators depending on three coefficients\cite{I} $X^{ij}_{n}$, $Y^{ij}_{n}$ and $Z^{ij}_{n}$, \emph{explicitly depending  on the interwiner of the node}.

The boundary state was defined\cite{scattering1} as a 
gaussian wave packet, centered on the values determined
by the background geometry $\mathbf q$ of the kind   
$
\Phi_{\mathbf q}({\mathbf j},{\mathbf i}) = C\ 
 e^{-\frac{1}{2j_0}(\delta {\mathbf I}A\delta {\mathbf I})
+  i \phi \cdot \delta {\mathbf I}},
$ 
where $\delta \mathbf{I}=(\delta {\mathbf{j}}, \delta {\mathbf{i}})$ is a 15d vector   with $\delta  {\mathbf{j}}$ and $\delta  {\mathbf{i}}$ given by the difference between the ten spins ${\mathbf{j}}$  and the five intertwiners ${\mathbf{i}}$ and their  background values $j_0$ and $i_0$ respectively.
$A$ is a $15\times15$ matrix and the normalization factor $C$ is determined by $\langle W |\Phi_{\mathbf q}\rangle=1$.  The spin phase coefficients are fixed by the background extrinsic geometry \cite{scattering1}. The intertwiner phase coefficients are fixed by requirement that the state remain peaked after a change of pairing to the value $i_0$. 
The crucial point is that the non commutativity \cite{tetraedro} of the the different angles of a tetrahedron, represented by the intertwiner variables in different pairings, requires a state with a phase dependence in the intertwiner variables to be peaked on the background angles in any pairing.
The correct value\cite{sc,I} for this is $exp\{i\frac{\pi}{2}i_n \}$.
The vertex and the state are written in terms of the intertwiner $i_n$, which is the virtual link of the node $n$ {\em in one chosen pairing}. 
It follows that  the vertex and the state  do not have 
the full symmetry of the 4-simplex and Eq.\eqref{partenza1} turn out not to be
invariant under $SO(4)$, as it should in the euclidean theory.
Two different strategies have been adopted in order to overcome this difficulty:  sum over the three pairings\cite{I} at each of the five nodes or choose an arbitrary pairing at each node and then symmetrize\cite{II} summing over the 5! permutations of the five vertices of the four-simplex. The first procedure don't allow to compute the correct propagator\cite{II}.

\section{Problems with the Barret Crane Vertex and Alternative Vertexes}
The last ingredient to compute the propagator is the vertex.
In Ref. \cite{scattering1} and Ref. \cite{I}, (a suitable adjustment of) the BC vertex was chosen
for $W$ and in this limit the propagator depends only on its asymptotic behavior, 
this has the structure \cite{asimp}
$
	W_{BC}(\mathbf{j})\sim
	e^{\frac{i}{2}(\delta{\mathbf{j}} G \delta {\mathbf{j}})} e^{i\Phi \cdot \delta{\mathbf{j}}} +e^{-\frac{i}{2}(\delta {\mathbf{j}}G \delta {\mathbf{j}})} e^{-i\Phi\cdot \delta {\mathbf{j}}}
	$
where $G$ is the $10\times10$ matrix given by the second derivatives of the 4d Regge action around
the symmetric state, and $\Phi$ is a 10d vector with all equal components, which were shown \cite{scattering1} to precisely match those determined by the background extrinsic curvature.
If we put together the three ingredients we end up with a sum of terms
\begin{equation}
	\tilde	{\mathbf G}_{{\mathbf q}\, n,m}^{\scriptscriptstyle ij,kl}	=	j_0^2 \sum{}_{ {\mathbf {j}}, {\mathbf {i}}} \;\overline{{W_{BC} }({\mathbf {j}})}  
K_n^{ij}K_m^{kl} \  \Phi({\mathbf j},{\mathbf i}),
\label{Gsigmandiagfine}
\end{equation}
(the tilde indicate that the component is not symmetrized under the 4-simplex symmetries)
where the $K_n^{ij}$ are linear expressions in $\delta i_n$ and $\delta j^{\scr{nm}}$ given in Ref. \cite{II}.
Computing this expression, the crucial point is that the phase in the link variable 
in the boundary state cancels with the phase of one of the two terms of $W_{BC}$, while 
the other term is suppressed \cite{scattering1} for large $j_0$ but \emph{the rapidly oscillating factor in the intertwiners variables is completely uncompensed by the dynamics and suppress the integral}\cite{I}. The intertwiner independence of the BC vertex prevents the propagator to have the correct long distance behavior. %Moreover there is a mismatch between the linear structures of $SO(4)$ (at the base of the BC vertex) and SO(3) (at the base of the LQG formalism used to build the operators and the boundary state) that prevents the propagator to have the correct tensorial structure. 

In Ref. \cite {II} has been proposed a vertex $W$ with an 
asymptotic form that includes a gaussian in all the 15 variables, and most \emph{crucially} a phase dependence also on the intertwiner variables. 
The proposed form for the asymptotic of $W$ is
	$W(\mathbf{j},\mathbf{i}) = 
	e^{\frac{i}{2}(\delta{\mathbf{I}} G \delta {\mathbf{I}})} e^{i\phi \cdot \delta{\mathbf{I}}} +e^{-\frac{i}{2}(\delta {\mathbf{I}} G \delta {\mathbf{I}})} e^{-i\phi\cdot \delta {\mathbf{I}}}
$  , where $G$ is a $15\times15$ real matrix, for which the only assumption is that it respects the symmetries of the problem and that it scales as $1/j_0$.  The quantity $\phi=(\phi_{nm},\phi_n)$ is now a 15d vector: 
its 10 spin components
$\phi_{nm}$ just reproduce the spin phase dependence of the BC vertex; while its five intertwiner components are equal and fixed to the value 
$\phi_n=\frac{\pi}{2}$.
This phase dependence is the crucial detail, that makes the calculation work because it allows the cancellation of the phases between the propagation kernel and the boundary state through which the dynamical kernel reproduces the 
semiclassical dynamics in quantum mechanics. If
this does not happens, the rapidly oscillating phases suppress the amplitude. 
In fact, now, all the phases 
in the boundary state cancels with the phase of one of the two terms of $W$, while 
the other term is suppressed for large $j_0$. Thus, \eqref{Gsigmandiagfine} reduces to 
$	\tilde	{\mathbf G}_{{\mathbf q}\, n,m}^{\scriptscriptstyle ij,kl}	= j_0^2 \sum{}_{{\mathbf {j}}, {\mathbf {i}}} 
	\;e^{-\frac{1}{2j_0}M_{\alpha\beta}\delta I_\alpha \delta I_\beta}
K_n^{ij}K_m^{kl} 
$
where $M=A+ij_0G$. 
This expression can be easily computed\cite{I} and the sum over permutations \cite{II} gives a propagator that can be matched with the perturbative one (in harmonic gauge, compatible with the radial\cite{elena claudio} gauge) fixing five free parameters in the boundary state.

The results of Ref. \cite{I} reinforce the idea that the BC model is not able to reproduce General Relativity (GR) in the low energy limit and have motivated the search for an alternative model \cite{EPR} able to reproduce GR.
Ref. \cite{II} shows that it is possible to recover the full propagator of the linearized theory from the LQG propagator and gives indications on the behavior that an alternative vertex can have to reproduce GR. In particular it requires for the new models an oscillation in the intertwiners that can be analyzed with analytical and numerical methods \cite{num}. Some preliminary numerical indications on one of the new models appear to show this dependence \cite{numeric}.

\end{document}